\documentclass[conference]{IEEEtran}
\IEEEoverridecommandlockouts
\usepackage{cite}
\usepackage{amsmath,amssymb,amsfonts}
\usepackage{algorithmic}
\usepackage{graphicx}
\usepackage{textcomp}
\usepackage{xcolor}
\usepackage{booktabs}
\usepackage{multirow}   
\usepackage{stfloats}
\def\BibTeX{{\rm B\kern-.05em{\sc i\kern-.025em b}\kern-.08em
    T\kern-.1667em\lower.7ex\hbox{E}\kern-.125emX}}
\begin{document}

\title{FedDetox: Robust Federated SLM Alignment via On-Device Data Sanitization\\

\thanks{This work was supported by JST SPRING, Grant Number JPMJSP2108.}
}

\author{\IEEEauthorblockN{Shunan Zhu}
\IEEEauthorblockA{
\textit{The University of Tokyo}\\
Tokyo, Japan \\
shunan-zhu@g.ecc.u-tokyo.ac.jp}
\and
\IEEEauthorblockN{Jiawei Chen}
\IEEEauthorblockA{\textit{The University of Tokyo}\\
Tokyo, Japan \\
jwchen@g.ecc.u-tokyo.ac.jp}
\and
\IEEEauthorblockN{Yonghao Yu}
\IEEEauthorblockA{\textit{The University of Tokyo}\\
Tokyo, Japan \\
y\_yu@cvm.t.u-tokyo.ac.jp}
\and
\IEEEauthorblockN{Hideya Ochiai}
\IEEEauthorblockA{\textit{The University of Tokyo}\\
Tokyo, Japan \\
ochiai@g.ecc.u-tokyo.ac.jp}
}

\maketitle

\begin{abstract}
As high quality public data becomes scarce, Federated Learning (FL) provides a vital pathway to leverage valuable private user data while preserving privacy. However, real-world client data often contains toxic or unsafe information. This leads to a critical issue we define as unintended data poisoning, which can severely damage the safety alignment of global models during federated alignment. To address this, we propose FedDetox, a robust framework tailored for Small Language Models (SLMs) on resource-constrained edge devices. We first employ knowledge distillation to transfer sophisticated safety alignment capabilities from large scale safety aligned teacher models into light weight student classifiers suitable for resource constrained edge devices. Specifically, during federated learning for human preference alignment, the edge client identifies unsafe samples at the source and replaces them with refusal templates, effectively transforming potential poisons into positive safety signals. Experiments demonstrate that our approach preserves model safety at a level comparable to centralized baselines without compromising general utility.

\end{abstract}


\section{Introduction}
The remarkable capabilities of contemporary Large Language Models (LLMs) are underpinned by scaling laws that necessitate vast quantities of high-quality training data. Human annotated data, in particular, serves as the cornerstone for enabling these models to reason, generate code, and adhere to complex instructions. However, the field is currently approaching a precipice where high-quality public data is nearing exhaustion, creating a significant bottleneck for future advancements \cite{10.5555/3692070.3694094}. In response to this scarcity, Federated Learning (FL) \cite{pmlr-v54-mcmahan17a} has emerged as a critical paradigm. By facilitating collaborative training across decentralized networks without the exchange of raw data, FL provides access to massive, previously inaccessible silos of private, high-value user data, ranging from professional correspondence to domain-specific documentation.

\begin{figure}[t]
  \includegraphics[width=0.9\columnwidth]{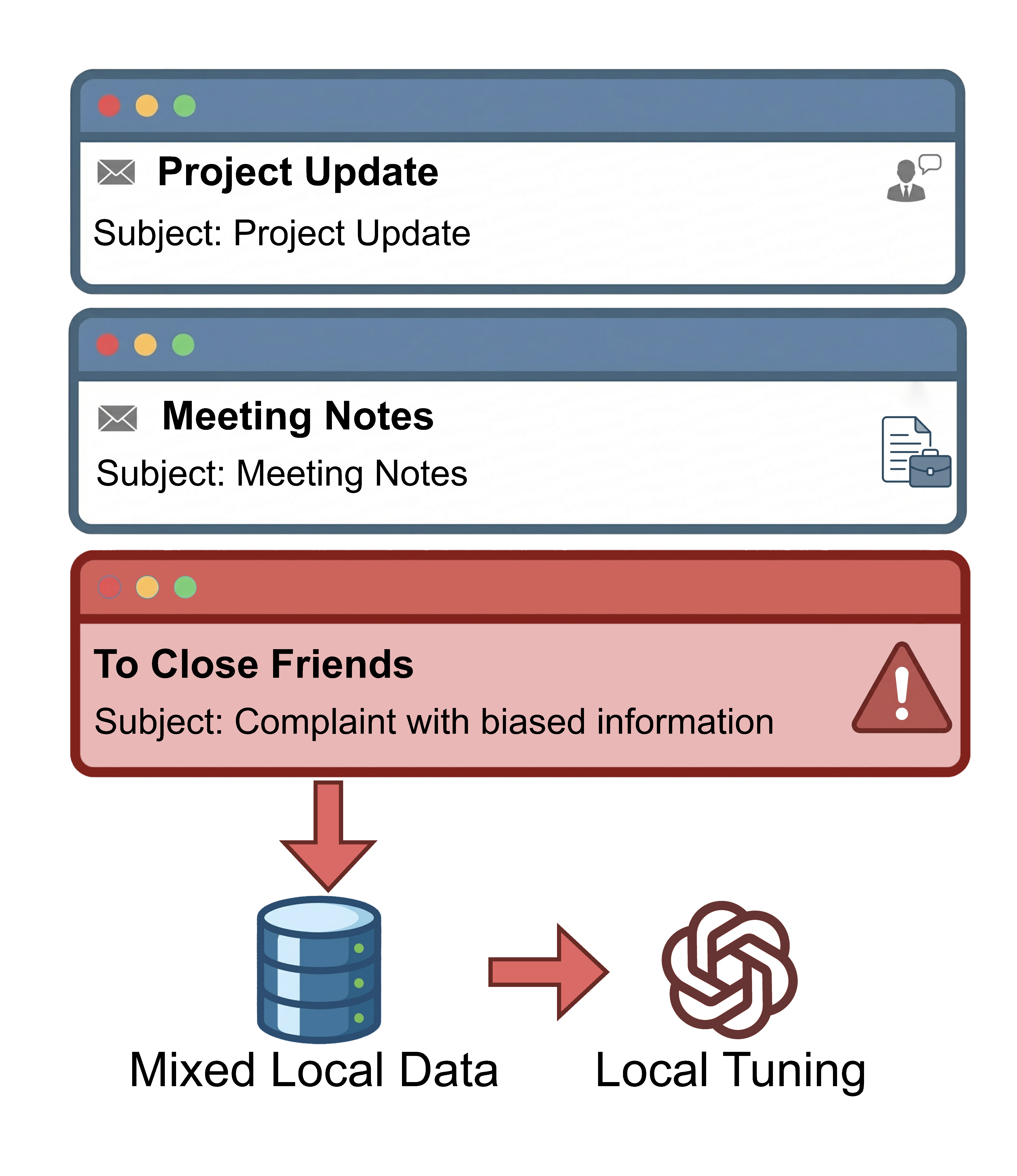}
  \caption{An illustrative example of unintended data poisoning, where daily emails containing sensitive emotions act as toxic samples during federated alignment.}
  \label{fig:email}
\end{figure}

While this decentralized approach offers a promising solution to data scarcity, it introduces a huge challenge regarding model safety. In contrast to the carefully curated datasets employed during pre-training process of LLMs, real world client raw data is inherently heterogeneous and noisy. Private datasets frequently contain toxic content, hate speech, or harmful instructions, not necessarily due to malicious intent, but often arising from the natural diversity of user interactions, such as a developer debugging vulnerable code or a user discussing sensitive sociopolitical topics. We characterize this phenomenon as \textbf{unintended data poisoning}. When global LLMs, especially Small Language Models (SLMs) \cite{vannguyen2024surveysmalllanguagemodels}, undergo federated preference alignment on such unsanitized data, the safety guardrails previously established in Reinforcement Learning from Human Feedback (RLHF) \cite{ouyang2022traininglanguagemodelsfollow} can be catastrophically compromised. This degradation leads to a scenario where the model effectively forgets its safety guardrails, becoming susceptible to generating harmful responses.

Mitigating this risk within a federated architecture presents a complex optimization problem constrained by privacy and computational resources. Safeguard models based on LLMs like Llama Guard \cite{inan2023llamaguardllmbasedinputoutput}, are too huge for deployment on edge devices with limited memory and computation power. Conversely, server side defenses typically operate on aggregated model updates rather than raw data, limiting their ability to pinpoint and exclude specific toxic samples without violating privacy protocols. Furthermore, existing lightweight filtering mechanisms often lack the semantic depth required to detect subtle adversarial prompts or context toxicity, forcing an undesirable trade off between the safety and the utility of the global model.

\begin{figure*}[!htbp]
\centering
\includegraphics[width=0.9\textwidth]{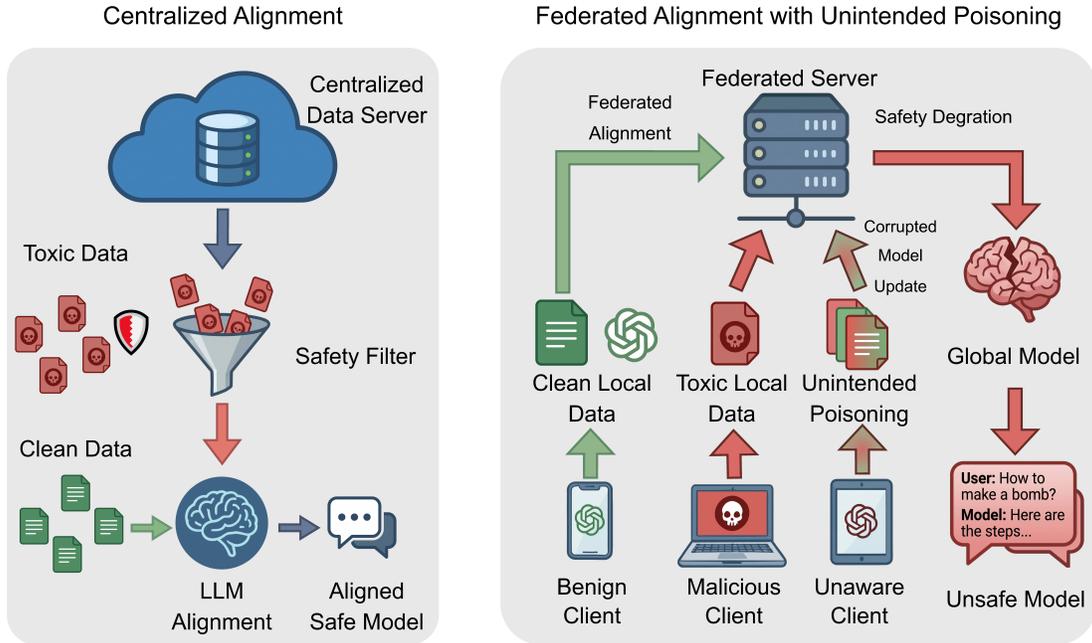}
\caption{\textbf{Problem Definition: The risk of Unintended Data Poisoning in Federated Alignment.} \textit{Left:} In centralized settings, raw data is aggregated and sanitized via a central \textbf{Safety Filter} before training, ensuring a safe model. \textit{Right:} In federated settings, raw data remains local and private. ``Unaware" clients, possessing a mixture of benign and toxic data like private chat history, perform local tuning without sanitization. This \textbf{Unintended Poisoning} propagates toxic gradients to the server, causing catastrophic \textbf{Safety Degradation} in the global model.}
\label{fig:def}
\end{figure*}

To bridge this gap, we propose \textbf{FedDetox}, \textit{\textbf{Fed}erated SLM Alignment \textbf{Detox}ify}, a novel framework for secure federated preference alignment that implements efficient, on-device data sanitization. Recognizing that edge devices cannot sustain the computational overhead of massive safety models, our approach employs knowledge distillation to transfer the sophisticated discernment capabilities of a large teacher model into a compact student classifier. To ensure this lightweight ``Guardian" remains robust against complex boundary cases, we create a training dataset covering as much distribution of the teacher model aspects. Specifically, during the federated training process, the Guardian functions as a local gatekeeper; rather than merely discarding unsafe data, it dynamically replaces toxic samples with safety-aligned refusal templates. This mechanism effectively transforms potential safety hazards into positive training signals, thereby reinforcing the model's ability to refuse harmful instructions without accessing raw user data.

This paper makes the following contributions to the field of trustworthy federated learning:
\begin{itemize}
    \item We formulate the problem of \textbf{unintended data poisoning} in federated SLM fine-tuning and identify the vulnerability of federated alignment to noisy client data.
    \item We propose \textbf{FedDetox}, a privacy-preserving framework that incorporates a comprehensive distillation pipeline. It compresses safety knowledge into lightweight client-side Guardians.
    \item Through extensive empirical analysis simulating federated environments with toxic private data, we demonstrate that our method effectively preserves the global model's safety alignment at a level comparable to centralized baselines, while maintaining high utility on downstream tasks and incurring negligible computational overhead.
\end{itemize}

\section{Related Work}

\subsection{Federated Alignment of Large Language Models}
The convergence of Federated Learning (FL) and LLMs has evolved from basic pre-training to instruction tuning, and recently, to human preference alignment. Early works in Federated Instruction Tuning (FedIT) demonstrated that aggregating diverse instructions from decentralized clients significantly enhances LLM generalization \cite{zhang2023building}. To mitigate the communication bottlenecks of full-parameter tuning, Parameter-Efficient Fine-Tuning (PEFT) methods, particularly Low-Rank Adaptation (LoRA), have become the standard in FL. For instance, FFA-LoRA \cite{sun2024improving} freezes randomly initialized adapters to reduce overhead, while LoRA-FAIR \cite{bian2024lora} addresses aggregation bias.

However, instruction tuning alone is not enough for ensuring model safety. Recent research has pivoted towards Federated Preference Alignment, adapting techniques like Reinforcement Learning from Human Feedback (RLHF) to federated settings. Frameworks such as PluralLLM \cite{srewa2025pluralllmpluralisticalignmentllms} attempt to align models with diverse user preferences while preserving privacy. More recently, Direct Preference Optimization (DPO) \cite{rafailov2023direct}, which optimizes the policy directly without a reward model, has gained traction in FL due to its stability and memory efficiency compared to PPO-based RLHF \cite{schulman2017proximalpolicyoptimizationalgorithms}. Our work builds upon this strategy, specifically focusing on the robustness of Federated DPO for resource-constrained Small Language Models (SLMs).

\subsection{Safety Vulnerabilities and Defenses in FL}
While FL preserves privacy, it is vulnerable to malicious updates. Traditional defenses focus on \textbf{\textit{Byzantine attacks}} that attempt to destroy model convergence \cite{10063647,blanchard2017byzantinetolerantmachinelearning}. However, these methods struggle against \textbf{\textit{Backdoor attacks}}, where adversaries inject stealthy semantic triggers \cite{NGUYEN2024107166,zhao2026revisitingbackdoorthreatfederated}. By maintaining statistical similarity to benign updates, backdoor attacks manipulate specific outcomes without degrading general utility, rendering them invisible to statistical outliers detection.

\subsection{Safety Guardrails in LLMs}

In the context of LLMs, the threat landscape shifts from gradient manipulation to \textit{Instruction Poisoning}. Recent work \cite{ye2024emerging} revealed that malicious clients can compromise global safety alignment by injecting unaligned instructions, a vulnerability that traditional robust aggregation algorithms fail to defend against. Furthermore, Pathmanathan \cite{pathmanathan2024poisoning} showed that even a small fraction (0.5\%) of poisoned data can corrupt the DPO process, leading to ``safety unlearning." Unlike prior works that focus on defending against active adversarial attacks, our research addresses a more pervasive yet subtle threat: the \textit{unintended propagation} of toxic content from ``unaware" clients. This necessitates granular, instance-level sanitization rather than coarse-grained client-level filtering.

Ensuring output safety is critical for LLM deployment. Centralized safeguards, such as OpenAI moderation API \cite{markov2023holistic} or Llama Guard Models \cite{inan2023llamaguardllmbasedinputoutput}, offer toxicity detection. However, these models possess huge sizes and unsuitable for federated edge devices due to prohibitive memory and latency costs. Conversely, lightweight filters based on keyword matching or perplexity often lack the semantic depth to detect subtle jailbreak attempts.

\section{Problem Formulation}

\subsection{Federated Alignment with LoRA}
We consider a synchronous federated learning system comprising a central server and a set of $N$ clients, indexed by $k \in \{1, \dots, N\}$. The global SLM is parameterized by $\theta \in \mathbb{R}^d$. To adhere to the resource constraints of edge devices, we adopt the Low-Rank Adaptation (LoRA) formulation. Specifically, for a pre-trained weight matrix $W_0 \in \mathbb{R}^{d_{in} \times d_{out}}$, the update is constrained to a low-rank decomposition: $W_0 + \Delta W = W_0 + BA$, where $B \in \mathbb{R}^{d_{in} \times r}$ and $A \in \mathbb{R}^{r \times d_{out}}$ are trainable low-rank matrices with $r \ll \min(d_{in}, d_{out})$.

In communication round $t$, the server distributes the global adapter parameters $\Phi^t = \{A^t, B^t\}$ to a selected subset of clients. Each client $k$ then performs local preference alignment on its private dataset $\mathcal{D}_k$ to minimize the loss. The server subsequently aggregates these local updates via weighted averaging: $\Phi^{t+1} = \sum_{k=1}^N p_k \Phi_k^{t+1}$.

\subsection{Threat Model: Unintended Data Poisoning}
In real-world deployment scenarios, assuming all client data is perfectly curated is unrealistic. We define the threat of \textit{Unintended Data Poisoning}, which fundamentally differs from malicious Byzantine attacks. We posit that the local dataset $\mathcal{D}_k$ is a mixture of two distinct distributions:
\begin{equation}
\mathcal{D}_k = \mathcal{D}_k^{clean} \cup \mathcal{D}_k^{toxic}
\end{equation}
Here, $\mathcal{D}_k^{clean}$ consists of benign instructions contributing to utility. In contrast, $\mathcal{D}_k^{toxic}$ comprises samples violating safety policies, such as hate speech or dangerous tutorials. Crucially, we assume clients are ``unaware" of this toxicity; these samples may originate from unfiltered web crawls, personal chat logs, or cached adversarial prompts.

Standard fine-tuning on $\mathcal{D}_k$ treats toxic samples as valid signals, minimizing the negative log-likelihood on $\mathcal{D}_k^{toxic}$. This effectively forces the model to unlearn its safety alignment, increasing the probability of generating harmful responses $y_{toxic}$ given a harmful prompt $x_{toxic}$:
\begin{equation}
P_{\Phi}(y_{toxic} | x_{toxic}) \uparrow
\end{equation}
This degradation propagates to the global model during aggregation, rendering the final SLM susceptible to jailbreaking despite initial alignment.

\subsection{Design Goals}
To counteract unintended poisoning without compromising privacy or utility, our \textbf{FedDetox} framework targets three objectives:
\begin{itemize}
    \item \textbf{Safety Preservation:} Significantly reduce the Attack Success Rate (ASR) of the fine-tuned global model on unseen toxic prompts.
    \item \textbf{Utility Maintenance:} Preserve the model's reasoning capability on benign benchmarks, avoiding performance degradation caused by aggressive filtering.
    \item \textbf{Edge Efficiency:} The sanitization module must be extremely lightweight ($|\phi_{student}| \ll |\theta_{SLM}|$) to ensure negligible latency and memory overhead on edge devices.
\end{itemize}

\section{Methodology}


\begin{figure*}[htbp]
\centering
\includegraphics[width=0.8\linewidth]{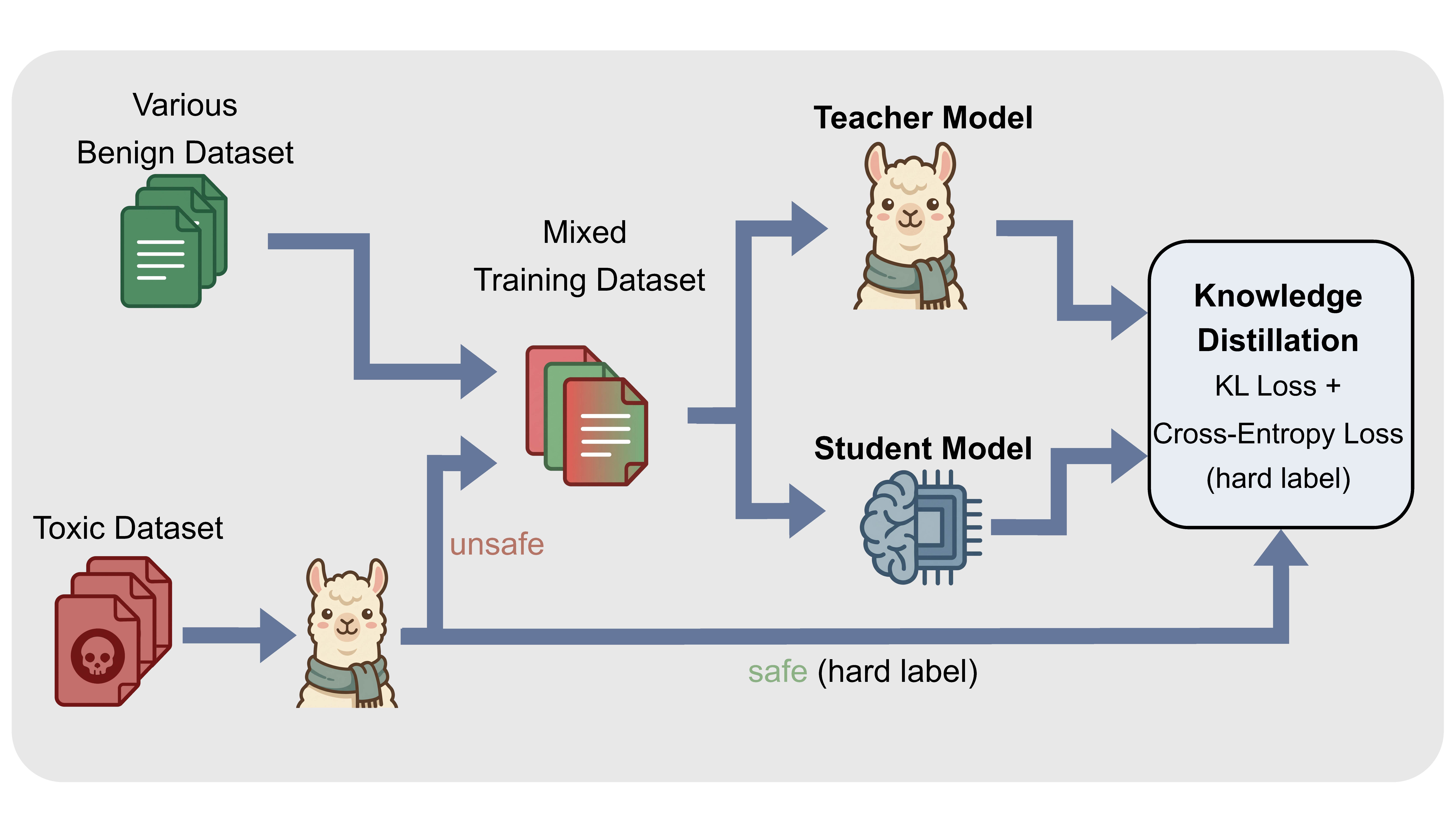}
\caption{Overview of the knowledge distillation pipeline for the lightweight Guardian. We construct a mixed dataset of benign and toxic samples to train the compact student model. The student learns to approximate the safety decision boundaries of the teacher by minimizing a joint loss function that combines the KL divergence of soft logits and the Cross-Entropy loss of hard labels.}
\label{fig:KD}
\end{figure*}

We propose \textbf{FedDetox}, a two-stage framework designed to address the conflict between the imperative for safety alignment and the computational constraints of edge devices in Federated Learning. The framework operates sequentially: \textbf{Phase I:} constructs a lightweight yet robust safety guardian via knowledge distillation, performed offline on the server side; \textbf{Phase II:} deploys this guardian to clients for real-time, privacy-preserving data sanitization during federated fine-tuning.

\subsection{Knowledge Distillation of the Lightweight Guardian}

The efficacy of FedDetox hinges on the capability of the client-side classifier to accurately detect toxicity with minimal computational overhead. We employ Knowledge Distillation (KD) to compress the safety capabilities of a large-scale teacher model into a compact student architecture. We utilize Llama Guard 3-8B as the teacher model ($\mathcal{T}$), leveraging its state-of-the-art semantic understanding and adherence to complex safety taxonomies. For the student model ($\mathcal{S}$), we select a compact architecture, MobileBERT, to ensure compatibility with the strict memory and latency budgets of edge devices. The objective is to train $\mathcal{S}$ to approximate the decision boundary of $\mathcal{T}$ such that for any input $x$, $P_{\mathcal{S}}(y|x) \approx P_{\mathcal{T}}(y|x)$, where $y \in \{\text{Safe}, \text{Unsafe}\}$. 
Besides, the teacher model sometimes also fails to identify unsafe data. We directly mark them as unsafe and use them as hard label in calculating loss function

\textbf{Loss Function.} The student model is trained on the augmented dataset $\mathcal{D}_{transfer}$. We minimize the Kullback-Leibler (KL) Divergence between the soft logits of the teacher and the student, allowing the student to learn both the binary label and the structural confidence of the teacher. The total loss function $\mathcal{L}_{KD}$ is defined as:
\begin{equation}
\begin{small}
    \mathcal{L}_{KD} = \alpha T^2 \cdot \text{KL} \left( \sigma \left( \frac{z_\mathcal{S}}{T} \right) \parallel \sigma \left( \frac{z_\mathcal{T}}{T} \right) \right) \\+ (1-\alpha) \cdot \mathcal{L}_{CE}(y_{true}, z_{\mathcal{S}})
\end{small}
\end{equation}
Where $z_{\mathcal{S}}$ and $z_{\mathcal{T}}$ are the logits of the student and teacher respectively, $T$ is the temperature parameter controlling the softening of probability distributions, $\sigma$ is the softmax function, and $\mathcal{L}_{CE}(y_{true}, z_{\mathcal{S}})$ represents the standard cross-entropy loss against ground truth labels.

\subsection{Federated Preference Alignment with On-Device Sanitization}

\begin{figure*}[htpb]
\normalsize
\begin{equation}
\mathcal{L}_{DPO}(\pi_\theta; \pi_{ref}) = -\mathbb{E}_{(x, y_w, y_l) \sim \mathcal{D}_k} \left[ \log \sigma \left( \beta \log \frac{\pi_\theta(y_w|x)}{\pi_{ref}(y_w|x)} - \beta \log \frac{\pi_\theta(y_l|x)}{\pi_{ref}(y_l|x)} \right) \right]
\label{eq:dpo}
\end{equation}
\hrulefill
\vspace*{4pt}
\end{figure*}

Once trained, the lightweight guardian $\phi_{\mathcal{S}}$ is distributed to all participating clients. In this phase, we perform \textbf{Federated Direct Preference Optimization (FedDPO)} directly on the local data on client based on SLMs.

Local Sanitization Mechanism.Upon receiving the global model, client $k$ initiates local training on its private dataset $\mathcal{D}_k$. Before a data sample $(x_i, y_i)$ enters the computation graph for gradient updates, it undergoes a forward pass through the local guardian. The guardian evaluates the prompt $x_i$ to determine a safety score. If the probability $P_{\phi_{\mathcal{S}}}(\text{Unsafe} \mid x_i)$ exceeds a pre-defined threshold $\tau$, the sample is flagged as toxic.

\textbf{Local Sanitization Mechanism.} Upon receiving the global model parameters, client $k$ initiates the local alignment process. Before a data sample enters the computation graph, it undergoes a forward pass through the local guardian. The guardian evaluates the prompt $x_i$ to determine a safety score. If the probability $P_{\phi_{\mathcal{S}}}(\text{Unsafe} \mid x_i)$ exceeds a pre-defined threshold $\tau$, the sample is flagged as toxic.

\textbf{Refusal Template Replacement.} A straightforward defense would be to discard flagged samples. However, merely removing toxic data creates a ``knowledge void," leaving the SLM vulnerable to similar prompts during inference. To convert this vulnerability into a defensive capability, FedDetox employs a \textit{Refusal Template Replacement} strategy. When a prompt $x_i$ is flagged as unsafe, we construct a synthetic preference pair to explicitly align the model against toxicity.

Specifically, we assign the safety-aligned refusal template $y_{refusal}$ as the chosen response $y_w$, and the original toxic response $y_i$ as the rejected response $y_l$. The local client then optimizes the DPO objective directly, see \ref{eq:dpo}:

where $\pi_\theta$ is the policy being trained, $\pi_{ref}$ is the frozen reference model, $\sigma$ is the sigmoid function, and $\beta$ controls the deviation from the reference. By minimizing this loss, the model learns to increase the likelihood of safe refusals while suppressing the probability of toxic outputs.

\textbf{Privacy and Efficiency Analysis.} Crucially, the entire sanitization process occurs strictly on-device. The raw private data $\mathcal{D}_k$ is never transmitted to the server; only the gradient updates derived from the sanitized data are shared. Furthermore, since the guardian is orders of magnitude smaller than the SLM backbone and requires no backward propagation, the computational overhead is negligible, preserving the efficiency required for edge deployment.

\section{Experiments}
We present a comprehensive evaluation to validate the effectiveness of FedDetox. Our experiments are designed to answer two fundamental questions: (1) \textbf{Impact of Poisoning:} How significantly does unintended data poisoning degrade the safety alignment of a global SLM in a federated setting? (2) \textbf{Efficacy of Defense:} Can our proposed on-device sanitization framework restore safety without compromising the SLM general utility?
\subsection{Experimental Setup}

To simulate a realistic federated learning scenario under edge resource constraints, we utilize \textit{Qwen2.5-1.5B-Instruct} as the global backbone \cite{qwen2025qwen25technicalreport}. This model strikes an optimal balance between parameter efficiency and instruction-following capability, making it a representative candidate mobile deployed SLMs. We implement the federated fine-tuning process using Direct Preference Optimization (DPO) \cite{rafailov2023direct}. Compared to other alignment methods, DPO is more effective for aligning models with safety preferences. We employ LoRA for parameter efficient updates (rank $r=8$, alpha $\alpha=16$). The federated environment consists of $N=10$ clients. In each round, $N=2$ clients are selected, we run 100 communication round for each settings. FedAvg is used as the central adapter aggregation method. Client model is tuned with a local batch size of 6 for 1 local epoch. We construct a composite dataset to simulate the mixture of benign and toxic data: \textit{Benign Data:} Sampled from the \texttt{distilabel-intel-orca-dpo-pairs} dataset \cite{mukherjee2023orca}, representing high-quality, safe user instructions. \textit{Unintended Poisoning Data:} Constructed using samples from \texttt{hh-rlhf} \cite{ganguli2022red} and \texttt{Sorrybench} \cite{xie2025sorrybench}. We flip the label of ``chosen" and ``reject" from \texttt{hh-rlhf} to identify them as harmful prompts. These diverse prompts including hate speech, violence, paired with compliant responses, simulating the ``unaware" toxic data on client devices. We collect around 16k prompts in total, 40\% of which is harmful. All experiments are conducted on NVIDIA GH200 Grace Hopper Superchips.

\subsection{Classifier Distillation and Deployment}

For the defense mechanism, we establish a robust Teacher-Student distillation pipeline: \textbf{Teacher Model:} Llama Guard 3-8B, a state-of-the-art safety classifier aligned with the MLCommons taxonomy \cite{dubey2024llama3herdmodels}. \textbf{Student Model:} MobileBERT (25M parameters) \cite{sun-etal-2020-mobilebert}, selected for its extreme lightness and low inference latency on edge devices.

\begin{figure}[!h]
\centering
\includegraphics[width=\columnwidth]{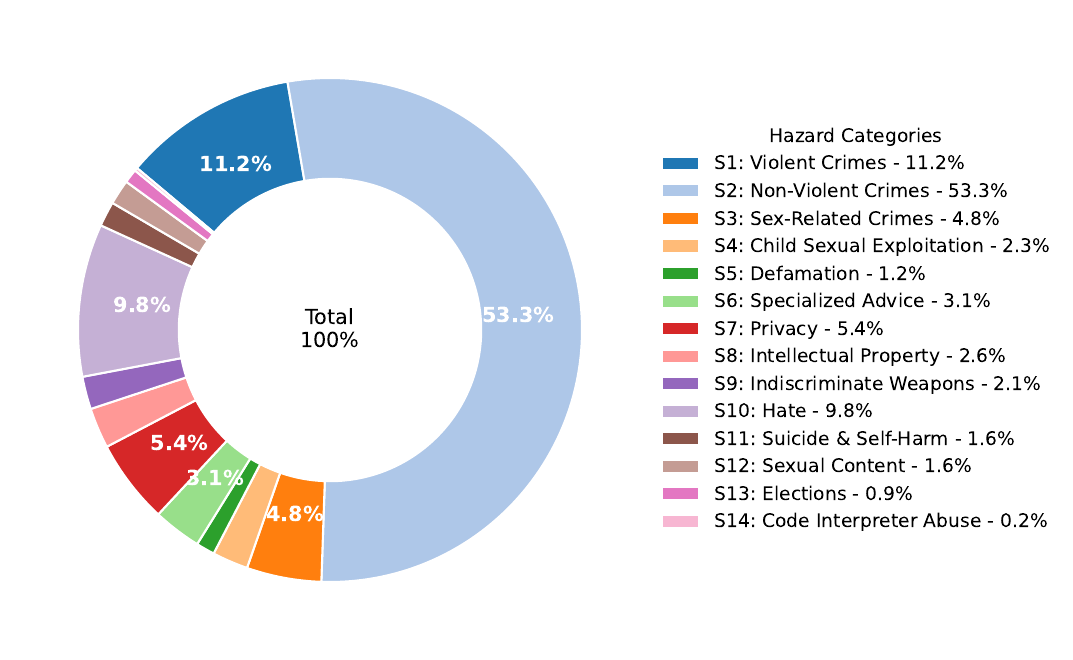}
\caption{Hazard category distribution of the distilled dataset. The dataset is a composite aggregation from multiple sources to ensuring comprehensive coverage of the MLCommons taxonomy.}
\label{fig:hazard_dist}
\end{figure}

We construct a specialized distillation dataset containing over 14k samples. The dataset is balanced with 60\% benign data and 40\% malicious data to ensure the Guardian learns distinct decision boundaries. Malicious prompts are collected from \texttt{PKU-SafeRLHF} \cite{ji2024pku}, \texttt{ToxicChat} \cite{lin2023toxicchat}, and a small set of handcrafted adversarial prompts. The distribution of hazard categories is shown in Fig. \ref{fig:hazard_dist}.

\subsection{Malicious Prompt and Jailbreak}

We evaluate the robustness of FedDetox against adversarial attacks. Our trained SLM demonstrates superior safety properties under three distinct categories of attacks, ranging from direct injections to sophisticated dynamic jailbreaks.
We utilize three key indicators to comprehensively assess the model's refusal capabilities: \textbf{Direct Malicious Prompts:} We measure the Attack Success Rate (ASR) on the AdvBench dataset \cite{zou2023universal}. This metric primarily evaluates the model's ability to refuse explicit malicious prompts and direct injection attempts without complex wrappers.
\textbf{Long-Context Jailbreaks:} We employ the dataset from DAN (Do Anything Now)  \cite{shen2024donowcharacterizingevaluating}, which contains 1405 jailbreak instances. This benchmark reflects the model's capability to resist jailbreak prompts embedded within contexts or role-playing scenarios, which traditionally confuse smaller models. \textbf{Dynamic Iterative Attacks:} We utilize the Tree of Attacks with Pruning (TAP) framework \cite{10.5555/3737916.3739868}, which dynamically optimizes jailbreak prompts to bypass defenses. We selected 100 harmful questions from AdvBench as targets for iteration. Notably, to adapt the jailbreak methodology to the computational capabilities of SLMs, we configured the TAP algorithm with a reduced iteration depth compared to standard LLM settings.
Table \ref{tab:jailbreak} presents the performance comparison. Under the ``Unintended Poison" setting, the global model becomes highly vulnerable, with the ASR on TAP surging to {77.0\%}, indicating that the model has learned to comply with sophisticated attacks. In contrast, FedDetox significantly fortifies the model. It maintains a low ASR of {14.0\%} on AdvBench and suppresses the TAP ASR to {61.0\%}, outperforming even the original Instruct model (67.0\%). This demonstrates that our Guardian-based sanitization effectively intercepts both static and dynamic attack vectors.

\begin{table}[htbp]\centering\caption{Robustness Evaluation against Jailbreak Attacks (ASR). Lower ASR indicates better safety alignment. Red represents worst performance, bold means the best.}\label{tab:jailbreak}
{\begin{tabular}{lccc}
\toprule
\textbf{Setting} & \textbf{Advbench} & \textbf{DAN Static} & \textbf{TAP} \\
\midrule
Original \textit{Qwen2.5-1.5B-Instruct} & \textbf{10.8\%} & \textcolor{red}{83.56\%} & 67.0\% \\
\midrule
FedDPO (Poisoned) & \textcolor{red}{30.8\%} & {78.51\%} & \textcolor{red}{77.0\%} \\
FedDetox (Ours) & {14.0\%} & \textbf{74.80\%} & \textbf{61.0\%} \\
\bottomrule
\end{tabular}}
\end{table}

\begin{figure}[htbp]
\centering
\includegraphics[width=\columnwidth]{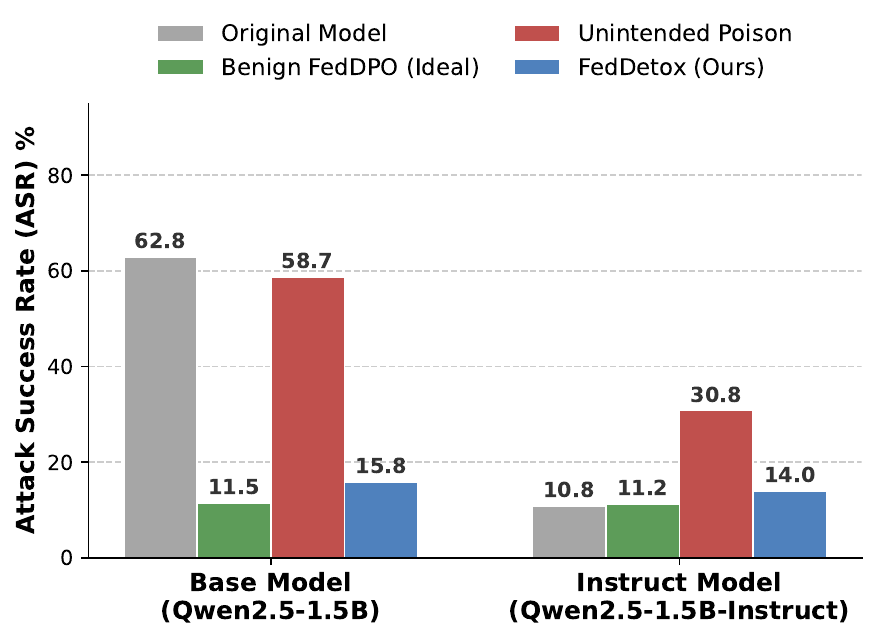}
\caption{Impact of poisoning and defense efficacy across models. \textbf{Gray bars} show the original model performance: the Base model is inherently unsafe (62.8\%), while the Instruct model is aligned (10.8\%). \textbf{Red bars} indicate that unintended poisoning causes the Base model to revert to high toxicity (\textcolor{red}{58.7\%}) and degrades the Instruct model. \textbf{Blue bars} demonstrate that FedDetox effectively suppresses toxicity in both scenarios, achieving safety levels comparable to the ideal benign setting (\textbf{Green bars}).}
\label{fig:asr_comparison}
\end{figure}

It is important to note that our global backbone, \textit{Qwen2.5-1.5B-Instruct}, has undergone rigorous safety alignment during the instruction tuning period, inherently possessing a low ASR (10.8\%). To isolate the contribution of our method from this pre-existing safety prior, we conducted an additional experiment using the raw \textit{Qwen2.5-1.5B-Base} model, which lacks inherent refusal capabilities. We add a benign only situation here simulating aligning the Base model solely with high-quality dpo pairs from \texttt{distilabel-intel-orca-dpo-pairs} \cite{mukherjee2023orca}. This yields a ``Ideal" model that follows instructions well but has learned safety boundaries from scratch, proving that DPO can align safety if data is clean. Figure \ref{fig:asr_comparison} provides a granular breakdown. The gray bars highlight the stark difference in initial safety between the Base and Instruct models. Under unintended poisoning, the Base model fails catastrophically, showing that without explicit refusal training, it easily learns toxic behaviors. FedDetox not only protects the Instruct model but, more importantly, constructs a robust safety boundary for the Base model from scratch, reducing its ASR to 15.8\%, closely mirroring the ideal performance achievable with strictly benign data shown in green bars.

A robust safety defense must not over-refuse benign queries. We use the XSTest benchmark \cite{rottger-etal-2024-xstest} to evaluate the trade-off between identifying unsafe prompts (Compliance $\downarrow$) and serving safe prompts (Compliance $\uparrow$). Table \ref{tab:xstest_results} details the performance. The Unintended Poison model exhibits a high compliance rate on unsafe prompts (\textcolor{red}{44.0\%}), indicating a failure of safety guardrails. In contrast, FedDetox reduces unsafe compliance to 25.0\%, closely matching the baseline (24.0\%), while maintaining a high compliance rate on safe prompts (94.4\%). This indicates that FedDetox does not simply reject all inputs but retains the semantic capability to distinguish between truly toxic commands and benign queries.

\begin{table*}[htbp]
\centering
\caption{Performance Comparison on XSTest Benchmark. We report the rate of Full Compliance. Lower is better for Unsafe Prompts (Safety), while higher is better for Safe Prompts (Utility).}
\label{tab:xstest_results}
\begin{tabular}{lccccc}
\toprule
\multirow{2}{*}{\textbf{Setting}} & \multicolumn{2}{c}{\textbf{Unsafe Prompts}} & & \multicolumn{2}{c}{\textbf{Safe Prompts}} \\
\cmidrule{2-3} \cmidrule{5-6}
 & \textbf{Compliance} $\downarrow$ & \textbf{Refusal} $\uparrow$ & & \textbf{Compliance} $\uparrow$ & \textbf{Refusal} $\downarrow$ \\
\midrule
Original \textit{(Qwen2.5-1.5B-Instruct)} & 24.0\% & 76.0\% & & 92.8\% & 7.2\% \\
\midrule
FedDPO (Poisoned) & \textcolor{red}{44.0\%} & \textcolor{red}{56.0\%} & & 97.2\% & 2.8\% \\
FedDetox (Ours) & 25.0\% & 75.0\% & & 94.4\% & 5.6\% \\
\bottomrule
\end{tabular}
\end{table*}

\begin{table}[h]\centering\caption{Comparison of SLM General Utility under different federated conditions.}\label{tab:utility_comparison}
{\begin{tabular}{lcccc}
\toprule

\textbf{Setting} & \textbf{TruthfulQA} & \textbf{MMLU} & \textbf{GSM8K} \\
\midrule
Original \textit{(Qwen2.5-1.5B)} & \textbf{46.9\%} & 59.6\% & \textbf{60.7\%}  \\
\midrule
FedDPO (Benign) & 45.8\% & \textbf{60.0\%} & 56.8\% \\
FedDPO (Poisoned) & \textcolor{red}{33.9\%} & \textcolor{red}{57.9\%} & \textcolor{red}{53.5\%} \\
FedDetox (Ours) & 38.9\% & 59.1\% & 55.4\% \\
\bottomrule

\end{tabular}}
\end{table}

A common concern with safety alignment is the ``alignment tax": the potential degradation of general reasoning capabilities \cite{lin-etal-2024-mitigating}. We evaluate this trade-off using MMLU \cite{hendryckstest2021}, GSM8K \cite{cobbe2021gsm8k}, and TruthfulQA\cite{lin-etal-2022-truthfulqa}.

Table \ref{tab:utility_comparison} presents the results. We could see that unintended poisoning severely impacts the reliability of tuned models. The TruthfulQA score drops precipitously from 45.8\% to 33.9\%. This suggests that ``unaware" toxic data often contains hallucinations or deceptive content that corrupts the factual alignment process. Meanwhile our method demonstrates minimal utility loss. On MMLU, FedDetox achieves {59.1\%}, which is statistically indistinguishable from the ideal Benign setting ({60.0\%}) and the original base model ({59.6\%}). Similarly, on GSM8K, our method maintains performance (55.4\%) comparable to the Benign baseline (56.8\%).

These results confirm that our \textit{Guardian classifier} and \textit{Refusal Template Replacement} strategies are highly precise. They surgically target toxic semantics without pruning the original knowledge learned in SLMs.

\subsection{Ablation Studies} 
\begin{figure}[htbp]
\centering
\includegraphics[width=0.9\columnwidth]{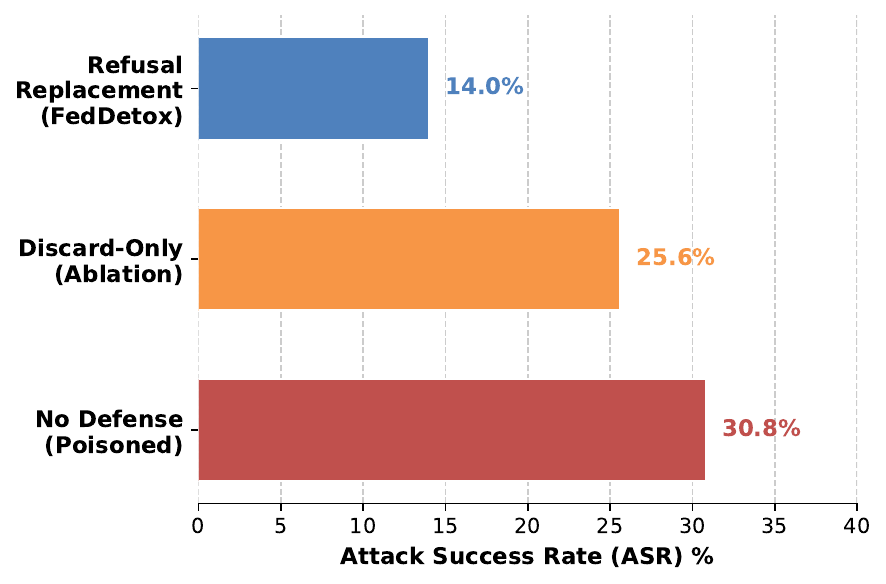}
\caption{Ablation Study: Effectiveness of sanitization strategies. Simply discarding toxic data (\textbf{Orange}) yields a significantly higher ASR (25.6\%) compared to FedDetox replacement strategy (\textbf{Blue}, 14.0\%). This confirms that explicit refusal training is necessary to define safety boundaries.}
\label{fig:ablation_discard}
\end{figure}

A critical component of FedDetox is the \textit{Refusal Template Replacement} strategy, which converts detected toxic queries into safety training signals. To validate the necessity of this mechanism, we compare it against a naive ``Discard-Only" baseline. In this ablation setting, any local data flagged by the Guardian is simply removed from the training set, preventing the model from seeing toxic samples but providing no explicit negative supervision. As shown in Figure \ref{fig:ablation_discard}, simply discarding toxic data results in a suboptimal safety alignment. The ASR on AdvBench rises to \textbf{25.6\%}, significantly higher than the \textbf{14.0\%} achieved by our replacement strategy.

This performance gap reveals a fundamental insight: merely shielding the model from toxic data creates a ``knowledge void." The model does not unlearn the potential toxicity inherent in its pre-trained weights, nor does it explicitly learn the boundary of \textit{what to refuse}. By contrast, our replacement strategy actively utilizes the toxic prompts as ``negative constraints" (the $y_l$ in DPO) paired with safe refusals ($y_w$). This forces the model to maximize the margin between compliant and refusal behaviors, effectively constructing a robust safety guardrail rather than just ignoring the threat.

\section{Conclusion}
In this work, we first formulated the critical problem of unintended data poisoning, identifying that the widespread presence of ``unaware" toxic data in real-world user interactions poses a severe and often overlooked threat to the safety alignment of federated models. To address this challenge, we proposed \textbf{FedDetox}, a robust framework tailored for Small Language Models (SLMs) on resource-constrained edge devices. By synergizing adversarial knowledge distillation with a novel Refusal Template Replacement strategy, FedDetox empowers clients to locally transform these potential safety hazards into explicit negative supervision signals. Extensive experiments demonstrate that our approach effectively restores safety guardrails against both static and dynamic jailbreaks to levels comparable with ideal benign baselines, all while incurring negligible computational overhead and preserving the general utility of SLMs.

\section*{Acknowledgment}
We thank all the anonymous reviewers for their careful reading of our manuscript and their insightful comments and suggestions.

\bibliography{references}
\bibliographystyle{IEEEtran}

\end{document}